\documentclass[a4paper,UKenglish,cleveref,autoref,thm-restate,nolines]{socg-lipics-v2021}

\usepackage{amsmath,amssymb,amsthm}
\usepackage{graphicx}
\usepackage{algorithm2e}
\usepackage{booktabs}
\usepackage{array}
\usepackage{xcolor}

\title{ Generating Diverse TSP Tours via a Combination of Graph Pointer Network and Dispersion}
\titlerunning{Generating Diverse TSP Tours via a Combination of Graph Pointer Network and Dispersion}

\author{Hao-Tsung Yang}{National Central University}{haotsungyang@gmail.com}{}{}
\author{Ssu-Yuan Lo}{National Central University}{lawrence090402@gmail.com}{}{}
\author{Kuan-Lun Chen}{National Central University}{glchen000000@gmail.com}{}{}
\author{Ching-Kai Wang}{National Central University}{a910424a@gmail.com}{}{}

\authorrunning{H.-H. Yang, S.-Y. Lo, K.-L. Chen, and C.-K. Wang}

\Copyright{}



\begin{document}

\maketitle

\nolinenumbers
\begin{abstract}
We address the Diverse Traveling Salesman Problem (D-TSP), a bi-criteria optimization challenge that seeks a set of $k$ distinct TSP tours. The objective requires every selected tour to have a length at most $c|T^*|$ (where $|T^*|$ is the optimal tour length) while minimizing the average Jaccard similarity across all tour pairs. This formulation is crucial for applications requiring both high solution quality and fault tolerance, such as logistics planning, robotics pathfinding or strategic patrolling. Current methods are limited: traditional heuristics, such as the Niching Memetic Algorithm (NMA) or bi-criteria optimization, incur high computational complexity ($O(n^3)$), while modern neural approaches (e.g., RF-MA3S) achieve limited diversity quality and rely on complex, external mechanisms (e.g., active search and relativization filters).

To overcome these limitations, we propose a novel hybrid framework that decomposes D-TSP into two efficient steps. First, we utilize a simple Graph Pointer Network (GPN), augmented with an approximated sequence entropy loss, to efficiently sample a large, diverse pool of high-quality tours. This simple modification effectively controls the quality-diversity trade-off without complex external mechanisms. Second, we apply a greedy algorithm that yields a 2-approximation for the dispersion problem to select the final $k$ maximally diverse tours from the generated pool. Our empirical results demonstrate state-of-the-art performance and efficiency. On the 52-city berlin instance, our average Jaccard index of $0.015$ significantly outperforms NMA ($0.081$) and the complex neural RF-MA3S ($0.509$). Furthermore, while achieving diversity quality comparable to complex bi-criteria approximation algorithms ($0.016$), our GPU-accelerated GPN structure maintains a superior empirical runtime growth close to $O(n)$. In a large instance with 783 cities, our solution is over 360 times faster than the bi-criteria alternative, delivering highly diverse, near-optimal TSP solutions with unprecedented speed and model simplicity.


\end{abstract}

\nolinenumbers
\section{Introduction}
In this work we address the Diverse Traveling Salesman Problem (D-TSP). Given an edge-weighted graph and a constant $c$, the objective of D-TSP is to identify a set $S$ of $k$ distinct TSP tours. This set must satisfy two competing criteria: first, the length of every tour $T \in S$ must be at most $c|T^*|$, where $|T^*|$ is the length of an optimal TSP tour; and second, the average Jaccard similarity among all pairs of tours in the set $S$ must be minimized. This formulation seeks high-quality solutions while maximizing their structural dissimilarity.

The concept of optimizing for solution diversity has attracted significant attention in recent years, driven by both theoretical advancements and practical application requirements. From a theoretical perspective, many classical combinatorial optimization problems have been revisited in their diverse forms~\cite{gao2022obtaining}, including minimum $s$-$t$ cuts ~\cite{de2023finding}, stable matchings ~\cite{deberg2025findingdiversesolutionscombinatorial}, and vertex cover~\cite{zhou2018efficient}. On the application side, the diversity objective serves as a potent heuristic method, offering substantial benefits across various domains. These benefits include effectively locating near-optimum solutions~\cite{qian2024quality,kim2021learning}, enhancing the robustness and stability of learned policies~\cite{yang2025towards,ganesh2025systemizing,pang2019improving,Jiang_Wu_Cao_Zhang_2022}, and providing crucial interpretability for opaque black-box models~\cite{mothilal2020explaining,dandl2020multi,chen2025pareto,nguyen2023feasible}.

\begin{figure}[htbp]
    \nolinenumbers 
    \centering
    
    \subfloat[Counter example tree-1]{%
        \includegraphics[width=0.31\linewidth]{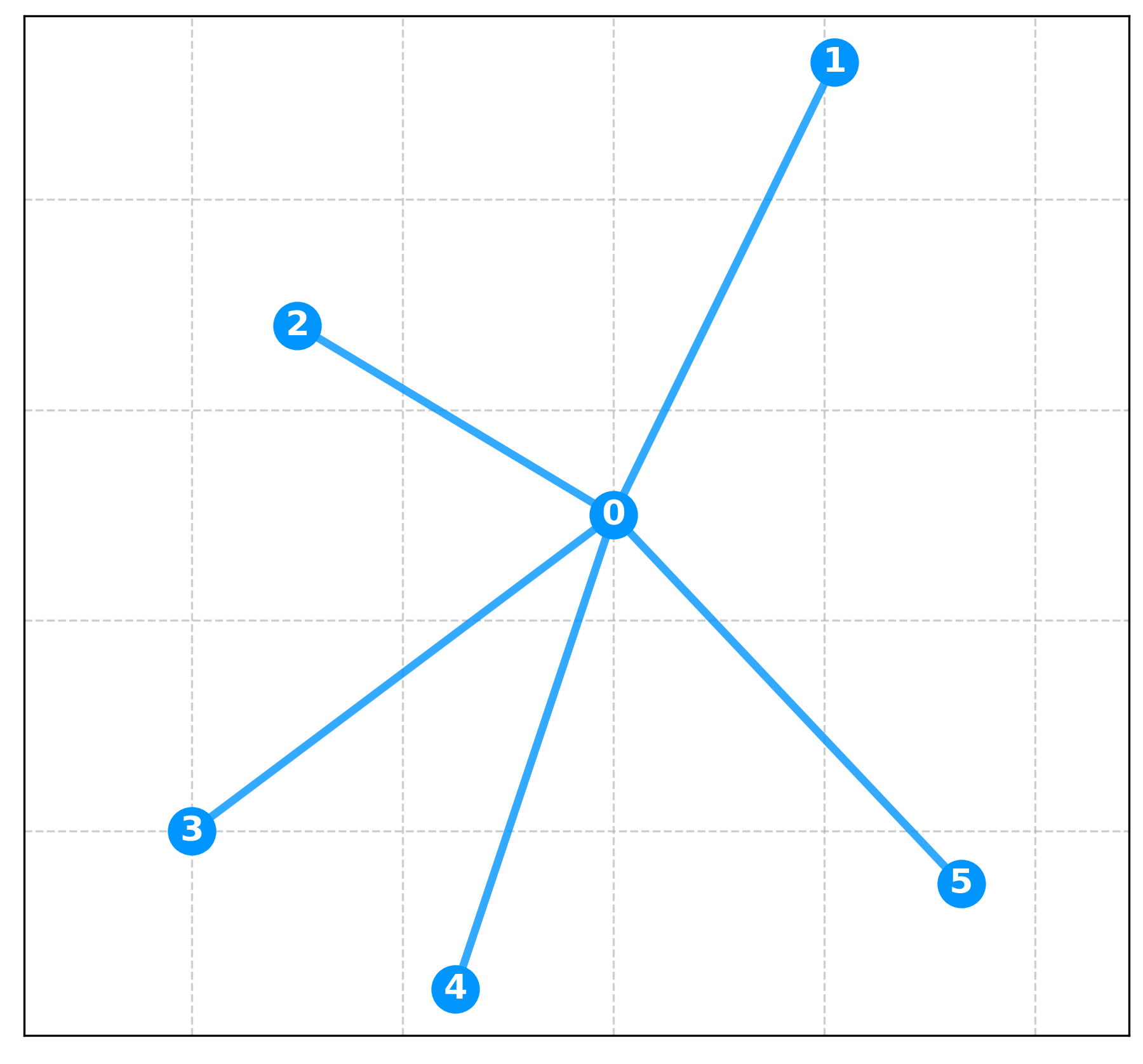}%
        \label{fig:counter_example_tree-1}%
    }
    \hfill 
    \subfloat[Counter example tree-2]{%
        \includegraphics[width=0.31\linewidth]{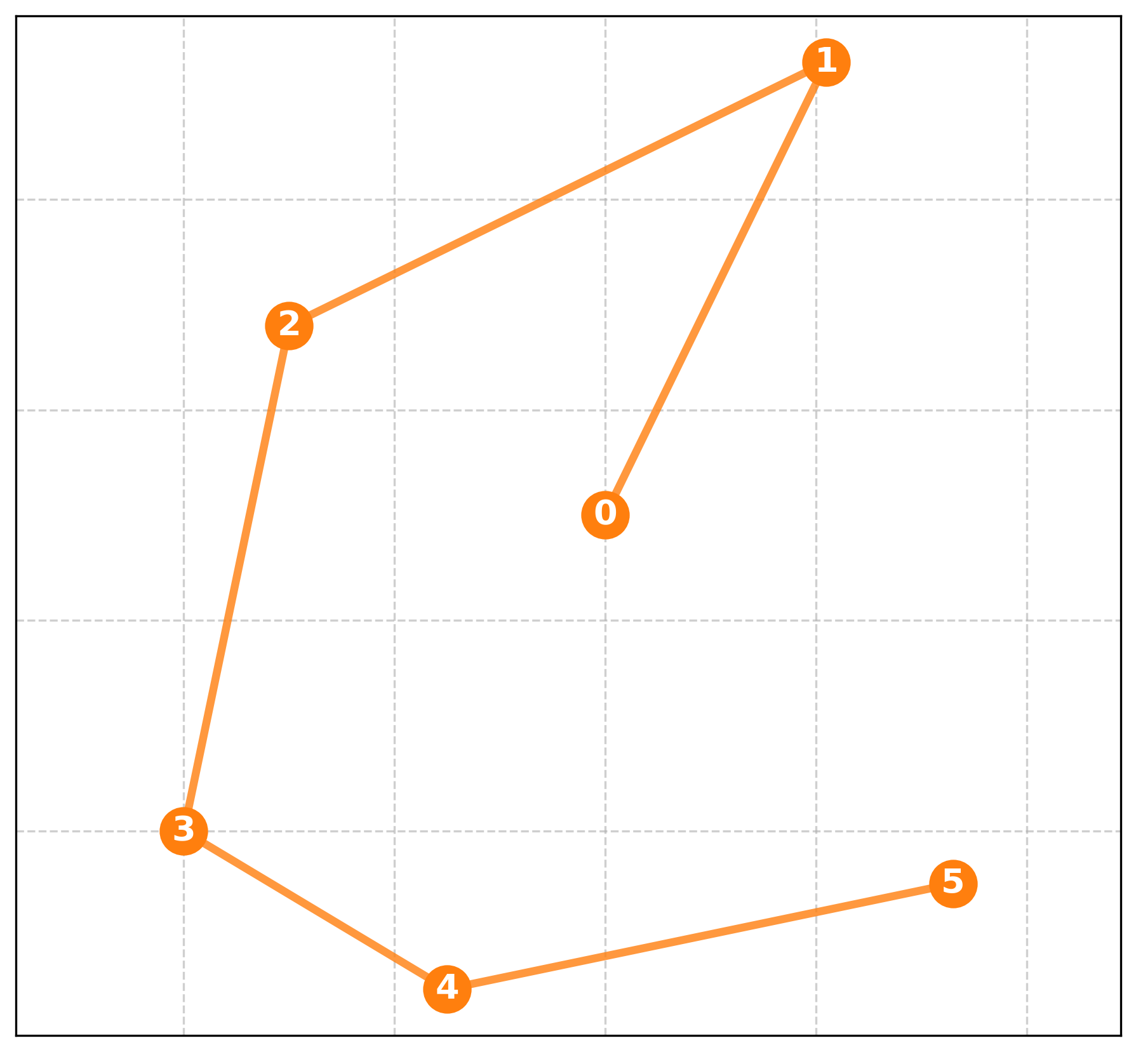}%
        \label{fig:counter_example_tree-2}%
    }
    \hfill 
    \subfloat[Tour]{%
        \includegraphics[width=0.31\linewidth]{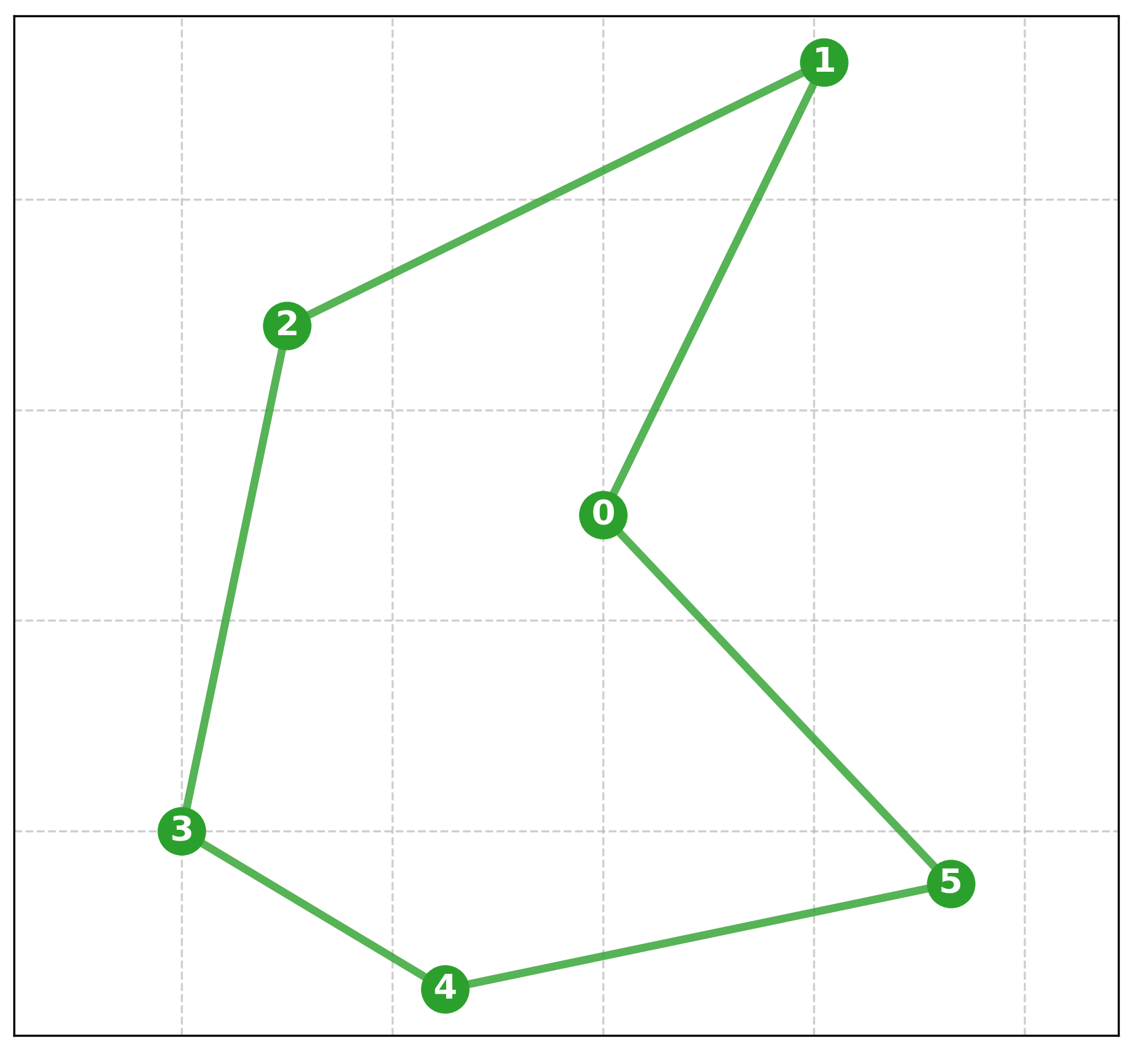}%
        \label{fig:counter_example_tour}%
    }
    
    \caption{A counterexample shows that one cannot preserve the dissimilarity when converting spanning trees into tours. (a) and (b) depict two distinct spanning trees sharing exactly one edge(resulting in a near-zero Jaccard index). Despite this minimal overlap, (c) demonstrates that distinct spanning trees can generate the exact same TSP tour via the double-tree heuristic.}
    \label{fig:counter_examples_combined}
\end{figure}

Within this growing theme, Diverse TSP also has many compelling potential applications. These include optimization tasks in logistics, where redundant and distinct path options are necessary for fault tolerance, and in robotics, such as efficient patrol planning and task allocation where various routes must be covered~\cite{yang2019patrol,liu2019task}. Despite this, the development of a guaranteed approximation algorithm for this problem is still open. The most closely related work are from the authors in~\cite{de2025price}, which focuses on 2 diverse tours only, and the authors in~\cite{gao2022obtaining}, which addresses the bi-criteria optimization of generating $k$ diverse approximate spanning trees. However, converting this approach as TSP tours suffers a critical limitation: the quality of the diversity achieved for spanning trees may fall apart when the trees are converted into Hamiltonian tours (see Figure \ref{fig:counter_examples_combined} as an example). Consequently, existing research addresses this challenge in heuristic ways, which can generally be divided into two main categories: traditional heuristics for Multi-Solution TSP (MSTSP) and neural approaches for Diverse TSP.

\textbf{Traditional Heuristics for MSTSP.} The Multi-Solution TSP has primarily been tackled by traditional heuristics~\cite{angus2006niching, do2022niching, han2018multimodal, huang2019niching, huang2018seeking, ronald1995finding}. Most of these methods are based on genetic or evolution-based algorithms, which naturally incorporate diverse "genes" or solution components during the evolution phase. The main difference among these lines of research is their focus on how to effectively explore and maintain the search space while simultaneously preserving the diversity and quality of the generated solutions. A significant drawback in this family of work is the high computation time required, as the problems scale up due to their NP-hard and multi-solution nature. Notably, the Niching Memetic Algorithm (NMA)~\cite{huang2019niching} has been identified as a state-of-the-art traditional heuristic for this problem~\cite{liu2023evolutionary}. NMA utilizes niche techniques that initially divide the generated TSP tours into distinct groups, maintaining tours only if they are similar to others within their assigned niche, thus protecting different regions of the solution space. Following NMA, approaches based on Evolutionary Diversity Optimization (EDO) have also been proposed~\cite{do2022niching}. EDO frameworks typically integrate diversity-specific metrics directly into the fitness function or use specialized evolution operators to push the population toward distinct areas of the solution space. Unlike NMA, EDO methods initialize from a known optimal solution and evolve the population solely to maximize diversity, which can be impractical given the NP-hard property of finding the optimal TSP tour.

\textbf{Neural Approaches for Diverse TSP.} In contrast to traditional heuristics, neural approaches for combinatorial optimization (CO) learn the solutions in a data-driven way. While these algorithms require a training stage, they offer the benefit of high-speed inference thanks to the parallel structure afforded by GPUs. The key challenge is developing policies that can effectively transfer to other graphs with different sizes and topologies. Most initial neural approaches focused on CO with single-criteria optimization, such as the foundational work of the Pointer Network~\cite{vinyals2015pointer, yin2020enhancing}, Graph Pointer Networks~\cite{ma2019combinatorial, yang2022graph}, and the Attention Model~\cite{kool2019attention, vaswani2017attention}. More recent work has shifted focus to multi-solution generation. By utilizing multiple parallel decoders, the model can explore different regions of the solution space~\cite{xin2021multi}. A key example is the work from the authors in~\cite{li2025diversity}, which proposes a deep reinforcement learning solver specifically for Diversity Optimization for TSP. This model is equipped with several mechanisms such as a relativization filter, temperature softmax~\cite{yi2019sampling}, and active search~\cite{hottung2022efficient}. These additional mechanisms, while effective, make the underlying model architecture complicated and tend to decrease overall interpretability. In addition, the active search method, while increases the quality of the solutions, slow down the inference stage significantly for large number of cities. 

\textbf{Our Contribution:} To address the D-TSP problem, we propose a novel framework that merges the strengths of traditional algorithms with a neural approach. Our solution breaks down the problem into two sequential steps. The first step is to efficiently sample a large pool of tours whose tour length is at most $c|T^*|$. The second step is to select $k$ maximally diverse tours from this pool. We show that the second step is mathematically related to the dispersion problem, where the greedy algorithm provides a guaranteed 2-approximation for this family of problems. The first step is relatively challenging: the sampled tours must be diverse to be representative of all high-quality tours less than $c|T^*|$, and the size of the pool must not grow exponentially. Thus, we utilize a graph pointer network model (GPN)~\cite{ma2019combinatorial} that incorporates a graph convolutional network~\cite{zhang2019graph} and the pointer network~\cite{vinyals2015pointer}. Surprisingly, we show that by simply adding a loss that approximate the sequence entropy in the decoding step, our model effectively controls the trade-off between the tour length and the diversity of the generated set. Compared with the work~\cite{li2025diversity}, our model is much simpler, operating successfully without any complex mechanisms and needless fine-tuning. 

The empirical studies show that our work significantly outperforms the state of the art with a few hours training on 40 cities only. For example, in the \textit{berlin} dataset (52 cities) with $k=60$ and $c=4$, the average Jaccard index (where lower is more diverse) of our method is $0.015$. This compares favorably against NMA~\cite{huang2019niching} at $0.081$ and the RF-MA3S~\cite{li2025diversity} at $0.509$, which is around 34 times higher. Our diversity performance is actually very close to the work in the approximated bi-criteria algorithm~\cite{gao2022obtaining}, which achieved $0.016$. However, our work is significantly faster since their approach involves solving a constraint programming problem with a total time complexity of $O(n^3)$. Although our GPN model theoretically takes $O(n^3)$ time complexity, by taking advantage of the fast parallel computing offered by the GPU structure, the empirical growth of the total running time is actually close to $O(n)$~\cite{blakely2021time, vaswani2017attention}. In a large instance like the \textit{rat} dataset (783 cities) when generating $1000$ tours, our solution takes approximately $1.2$ hours. This is about $60$ times longer than the 52-city case. In stark contrast, the bi-criteria algorithm takes around $1.8$ days, which is $360$ times longer, highlighting the superior efficiency of our neural-heuristic hybrid approach.

\section{Problem Definition}

We formulate the problem of generating diverse TSP solutions within the framework of the Multi-Solution Traveling Salesman Problem~\cite{li2025diversity}. Our objective is to identify a set of Hamiltonian cycles that are not only high-quality (near-optimal) but also distinct from one another.

\subsection{The Traveling Salesman Problem}
Let $G = (V, E, w)$ be a complete weighted graph, where $V = \{v_1, \dots, v_n\}$ is the set of $n$ vertices, $E$ is the set of edges, and $w: E \to \mathbb{R}^+$ is a weight function representing the Euclidean distance between vertices. A solution to the TSP is a Hamiltonian cycle (tour) $\pi$, defined as a permutation of $V$ or equivalently as a subset of edges $E_\pi \subset E$ such that every vertex has a degree of exactly 2 and the subgraph is connected. The cost of a tour is given by $C(\pi) = \sum_{e \in E_\pi} w(e)$. The objective of the standard TSP is to find a tour $\pi^*$ that minimizes this cost: $C(\pi^*) = \min_{\pi} C(\pi)$.

\subsection{Diverse TSP Formulation}
Different from the standard TSP which seeks a single optimum, the Diverse TSP aims to find a set of $k$ distinct tours, $\Pi = \{\pi_1, \pi_2, \dots, \pi_k\}$. This set is optimized according to two competing criteria: \textit{Solution Quality} and \textit{Diversity}.

\textbf{Quality Constraint.} To ensure the generated solutions are practically viable (e.g., for efficient logistics or security patrols), we impose a cost constraint relative to the optimal solution. A tour $\pi$ is considered valid only if its cost is within a factor $c \ge 1$ of the optimal cost:
\begin{equation}
    C(\pi) \le c \cdot C(\pi^*)
\end{equation}
where $c$ acts as a tolerance parameter controlling the allowable optimality gap. This defines the feasible search space of high-quality solutions, denoted as $\Omega_c$.

\textbf{Diversity Metric.} We quantify the diversity between any two tours $\pi_i$ and $\pi_j$ based on the dissimilarity of their edge sets. We utilize the Jaccard Index to measure similarity:
\begin{equation}
    J(\pi_i, \pi_j) = \frac{|E_{\pi_i} \cap E_{\pi_j}|}{|E_{\pi_i} \cup E_{\pi_j}|}
\end{equation}
Here, a lower Jaccard Index indicates higher diversity.

\textbf{Objective.} The goal of the Diverse TSP is to select a subset of $k$ solutions $\Pi \subseteq \Omega_c$ that minimizes the average pairwise similarity, thereby maximizing the dispersion of the solution set:
\begin{equation}
\label{eq:obj of max diversity}
    \min_{\Pi \subseteq \Omega_c, |\Pi|=k} \frac{1}{k(k-1)} \sum_{\pi_i \in \Pi} \sum_{\pi_j \in \Pi, j \ne i} J(\pi_i, \pi_j)
\end{equation}

\section{Method}

\begin{figure}
    \centering
    \includegraphics[width=1\linewidth]{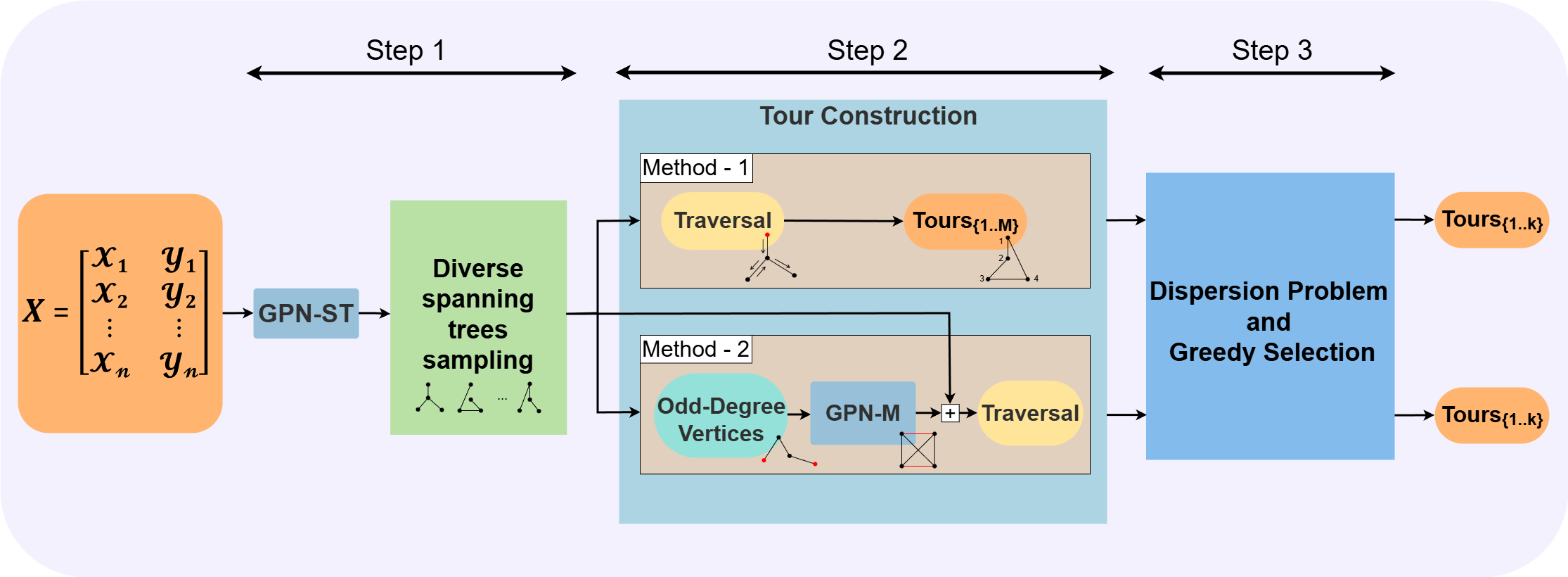}
    \caption{The proposed architecture integrating GPN for spanning tree and perfect matching generation, respectively, followed by a dispersion module for diverse tour selection.}
    \label{fig:flow}
\end{figure}

We propose a unified learning-based framework to generate a set of high-quality and diverse TSP tours. As illustrated in Figure \ref{fig:flow}, our approach operates in a "generate-then-select" manner, consisting of three sequential stages:

\begin{enumerate}
    \item \textbf{Diverse Spanning Tree Sampling:}
    The input graph is first processed by our GPN-ST (detailed in Section \ref{sec:model}) to sample a large pool of diverse spanning trees. These sampled trees serve as the geometric backbone for our solutions.

    \item \textbf{Tour Construction:}
    The sampled trees are transformed into valid TSP tours $\mathcal{T}_{valid}$ via two parallel inference strategies: \textit{Method 1} uses a classic double-edge heuristic with randomness sampling, while \textit{Method 2} adopts Christofides Algorithm but replace the perfect matching to a deep reinforcement learning approach that can generate diverse matching. In this work, this approach uses the network structure (GPN-ST) same as the step one.
    
    \item \textbf{Dispersion Problem and Greedy Selection:}
    Given the set of sampled tours $\mathcal{T}_{valid}$, the problem becomes how to select a subset of $k$ tours that maximizing the diversity (Equation~\ref{eq:obj of max diversity}). This problem is actually a reduction of Dispersion problem~\cite{ravi1994heuristic} which is NP-hard. Thus, we use a greedy method that incrementally select $k$ tours to maximize the diversity from $\mathcal{T}_{valid}$. According to recent work, this greedy manner provides 2-approximation which is also a tight bound assuming the exponential time hypothesis~\cite{gao2022obtaining}.
    
\end{enumerate}
The following are the details for step 2 and 3.

\subsection{Method 1: Diversity via Traversal}
\label{sec:method1}

The first strategy simply uses the double-tree heuristic to transform the sampled spanning trees into valid TSP tours. However, instead of directly traversing the double-tree, we use the uniform sampling for all the neighbor vertices while traversing to decrease the chance of generating same vertex order from different structures of the spanning trees (the example shown in Figure~\ref{fig:counter_example_tour}). That is, for each sampled spanning tree with doubling each edge, we perform a randomized Depth-First Search (DFS) which, at each step of the recursion, identifies all unvisited neighbors of the current vertex and selects the next vertex to visit by \textit{uniformly sampling} from this set. By maintaining a visited set during the traversal, we form a valid Hamiltonian cycle directly.

    

\subsection{Method 2: Christofides Algorithm with Diverse Matching}
\label{sec:method2}

The second method adapts the Christofides algorithm~\cite{christofides1976worst} but replaces the minimum weighted perfect matching to a deep reinforcement learning mechanism. This provides the benefits since the Christofides algorithm is deterministic and computationally intensive due to the matching step. Our approach introduces a deep matching network which using the same structure of GPN but the objective is changed to learn the bi-criteria matching with diversity and high quality. 

Given a sampled spanning tree $T$, we first identify the set of vertices $O \subseteq V$ that have an odd degree in the tree $T$ ($|O|$ is always even). Then, taking $O$ as an input, the deep matching network outputs a set of diverse matchings for the subgraph induced by $O$. Lastly, each matching is combined with $T$ to form an Eulerian multigraph. We then compute an Eulerian circuit on the multigraph and apply same procedure mentioned in Method 1 (DFS and uniform sampling) to produce the final TSP tour. Compared with the Christofides algorithm, our algorithm maintains the same  time complexity of $\mathcal{O}(n^3)$, where the $\mathcal{O}(n^3)$ is dominiated by the time of accessing the GCN structure in the network.

    


\subsection{Greedy Selection}
\label{sec:selection}

Given a pool of sampled tours $\mathcal{T}_{valid}$. We incrementally add the tour with a procedure we called the \textit{furthest insertion}. That is, starting from the empty set $S_0=\emptyset$, the $i$-th tour $\pi_i$ is added into $S_{i-1}$ where $\pi_i$ satisfies the following
\begin{equation}
\label{eq:greedy selection}
\begin{aligned}
\min_{\pi_i \in \mathcal{T}_{valid}} \quad 
    & \sum_{\pi_j \in \Pi_{i-1}} J(\pi_i, \pi_j) \\
\text{s.t.} \quad
    & \pi_i \notin \Pi_{i-1}, \\
    & C(\pi_i) \le c \cdot C(\pi^*).
\end{aligned}  
\end{equation}

Denote the size of $\mathcal{T}_{valid}$ as $M$. Naively solving Equation~\ref{eq:greedy selection} at iterative step $i+1$ takes $\mathcal{O}((M-i)i)$ time to calculate the pairwise similarity between two sets $S_i$ and $\mathcal{T}_{valid} \backslash S_i$. However, one can efficiently reduce the time to $\mathcal{O}(M-i)$ by recording on edge usage times among all tours in $S_i$.

The process begins by initializing the set $S$ with a randomly selected tour $\pi_1$ from $\mathcal{T}_{valid}$. We then maintain a global frequency map, $F$, which records the count of each edge $e$ appearing in the currently selected tours in $S$.

For each subsequent step $i = 2, \dots, k$, we select the next tour $\pi_i$ that minimizes the overlap with the current set $S_{i-1}$. Specifically, we define a penalty score for each candidate tour $\pi \in \mathcal{T}_{valid} \setminus S_{i-1}$ as the sum of the frequencies of its edges in $F$:
\begin{equation}
    \text{Score}(\pi) = \sum_{e \in E_{\pi}} F(e)
\end{equation}
where $F(e)$ denotes the number of times edge $e$ has appeared in the tours already added to $S$. One can see that the score of a tour is actually proportional to the jaccard index among all the tours in $S_{i-1}$. Thus, solving Equation~\ref{eq:greedy selection} is equivalent to selecting tour $\pi_i$

\begin{equation}
    \pi_i = \mathop{\arg\min}_{\pi \in \mathcal{T}_{valid} \setminus S} \text{Score}(\pi)
\end{equation}
Upon selection, we update $F$ by incrementing the counts for all edges in $\pi_i$. 

\noindent \textbf{Time Complexity:} 
Generating the entire pool of $M$ diverse spanning trees (Step 1) entails a total computational cost of 
$\mathcal{O}(n^3|S|)$. This is because each individual sampling operation requires $\mathcal{O}(n^2)$ time, largely dominated by the GPN-ST encoder's message passing on the fully connected graph and the decoder's sequential attention mechanism, both of which scale quadratically with the number of cities $n$. Step 3 is an iterative process. In $i-$th iteration, each tour in $S$ compares the dissimilarity among $i-1$-th tours. The total running time is at most $\mathcal{O}(k|S|)$. Thus, the overall complexity is $\mathcal{O}((n^2+k)|S|)$.

\section{Model}\label{sec:model}

\begin{figure}
    \centering
    \includegraphics[width=1\linewidth]{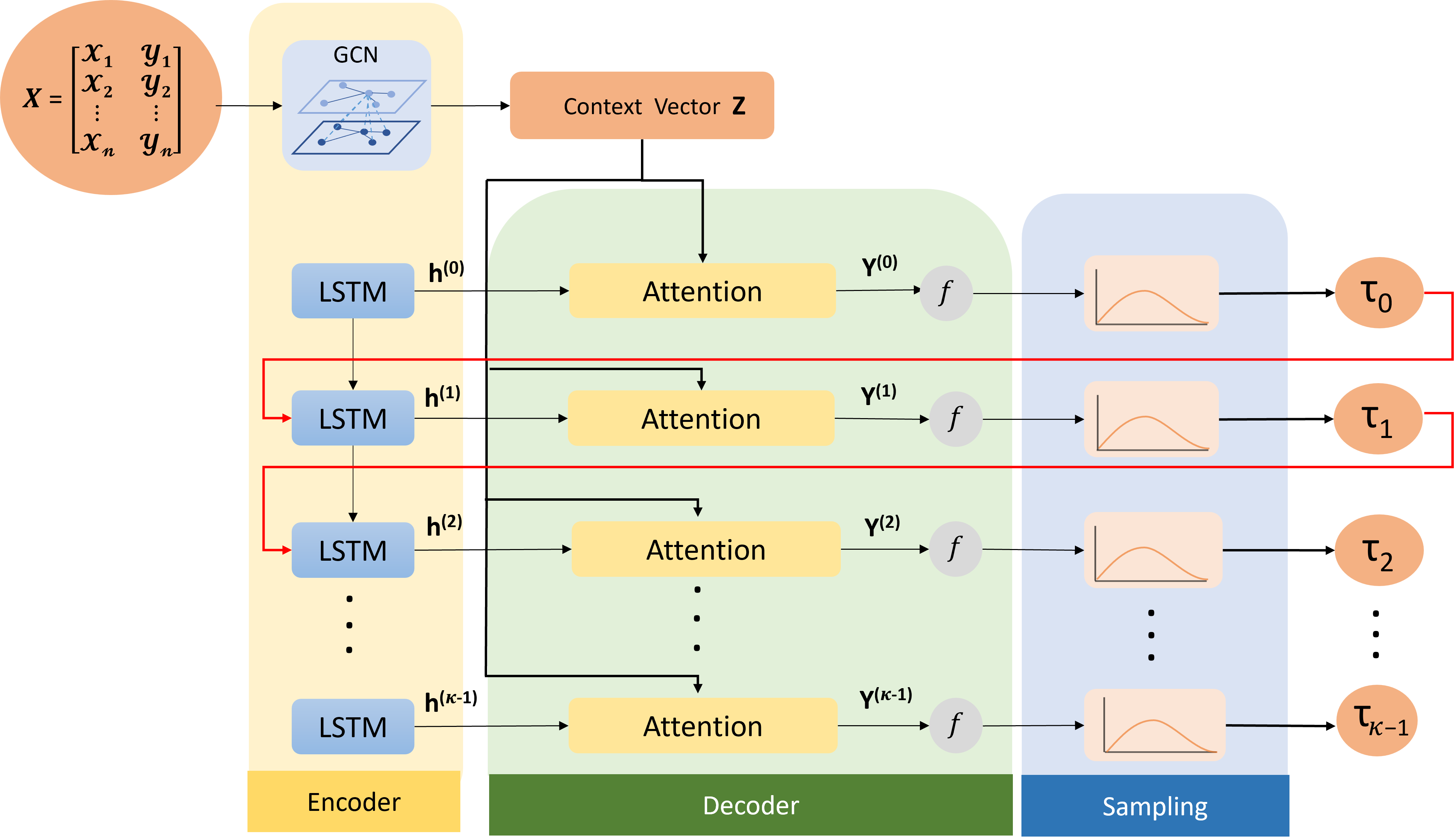}
    \caption{Overview of the model architecture.}
    \label{fig:model}
\end{figure}

We address both the spanning tree and the perfect matching as sequential edge-selection problems on Euclidean graphs. Let
\[
\mathbf{X} =
\begin{bmatrix}
x_1 & y_1 \\
x_2 & y_2 \\
\vdots & \vdots \\
x_n & y_n
\end{bmatrix}
\in \mathbb{R}^{n \times 2}
\]
denote the matrix of vertex coordinates, and let \(V=\{1,\dots,n\}\).
We define the complete weighted graph \(G=(V,E,w)\), where
\[
E = \{(i,j): i<j\}, 
\qquad 
w(i,j) = \|\mathbf{x}_i - \mathbf{x}_j\|_2 .
\]

The construction of the graph solution is modeled as a sequential edge selection processed within Markov decision process(MDP). A tour solution is an ordered list of distinct edges $\tau = (\tau_0, \tau_1, ..., \tau_{\kappa-1})$. For a spanning tree, the sequence length of $\kappa$ is $n-1$, and for a perfect matching, $\kappa$ is $\frac{n}{2}$. At each step t, the state is the sequence of previously chosen edges $\tau_{0:t-1}$. The action space consists of the remaining edges that keep the partial solution feasible when added, for spanning trees, edges that do not create a cycle; for perfect matching, edges whose endpoints are currently unmatched. To align with the optimization goals, we define the cumulative reward for a complete sequence as the negative total edge weight, encouraging the policy to minimize the objective. We define a sequence policy $p(\tau \rvert G)$, parameterized by $\theta$, as an sequential distribution over feasible actions:

\begin{equation}
p(\tau \mid G) = \prod_{t=0}^{\kappa-1} p_{\theta}\!\big(\tau_t \mid G, \tau_{0:t-1}\big)
\label{eq:autoregressive-policy}
\end{equation}
where $\tau_t$ is the edge selected in step t and $\tau_{0:t-1}$ is the partial solution, and $G$ is the input graph instance. Each conditional $p_{\theta}\!\big(\tau_t \mid G, \tau_{0:t-1}\big)$ places probability only in the feasible action set. 
The model framework is depicted in Figure \ref{fig:model}. The GCN transforms the graph into invariant edge embeddings, and the LSTM compresses the sequence of previously selected edges into a hidden state. The attention mechanism then integrates these embeddings to compute the selection probabilities and determine the next edge. The LSTM and attention mechanism operate sequentially, repeating until the entire solution is generated. Our model architecture is based on the work in~\cite{ma2019combinatorial}.

\textbf{Encoder: }
The encoder integrates a Graph Convolutional Network(GCN)-style message passing scheme and a Long Short-Term Memory(LSTM) network. The GCN is responsible for context generation: the vertex  coordinates $x_v$ are first embedded, and iteratively update over $L$ layers via a residual message passing scheme to capture the graph's global structural context. The update of $l$-th. ($l \leq L$) layer can be described as : 

\begin{equation}
    x_u^l=\gamma x_u^{l-1}\Theta+(1-\gamma)\phi_\theta\left (\frac{1}{\left|\mathcal{N} (u)\right|}\left \{x_v^{l-1} \right\}_{v\in\mathcal{N}(u)\cup \{u\}}  \right ) 
\end{equation}
Here, $\phi_\theta : \mathbb{R}^{d_{l-1}} \rightarrow \mathbb{R}^{d_l}$ denotes the aggregation function. The parameter $\gamma$ is the proportion of the previous feature of the vertex $x_u^{l-1}$, which is first transformed by the learned weight matrix $\Theta$. The remaining proportion, $(1-\gamma)$, is dedicated to the aggregated information from the vertex's neighbors $\mathcal{N}(u)$ and itself. 
The final Context $\textbf{Z}$ is derived as a set of edge embeddings $\textbf{z}_{uv}$, which are constructed by summing the initial edge embedding $\eta_{uv}$ with the final updated embeddings of its two endpoints $x_{u}^{(L)}$ and $x_{v}^{(L)}$, where $\textbf{z}_{uv} = \eta_{uv} + x_{u}^{(L)} + x_{v}^{(L)}$. Concurrently, the LSTM network handles dynamic sequential state tracking. In subsequent steps $t$, the LSTM updates its hidden state ${h}^{(t)}$ and ceil state ${c}^{(t)}$ by processing the embedding $\textbf{z}_{uv}$ of the previously selected edge (i.e., $(u,v)=\tau_{t-1}$) and outputs the dynamic Query Vector ${q}^{(t)}$. Both this dynamic ${q}^{(t)}$ and $\textbf{Z}$ are then passed to the Decoder's Attention module to determine the next action.

\textbf{Decoder: }
The Decoder Module is built upon an Attention mechanism and supported by edge masking logic. At each step $t$, it utilizes the dynamic Query Vector $q^{(t)}$ and the GCN's Context Vector $\textbf{z}_{uv}$ to compute compatibility scores $Y_{uv}^{(t)}$ for edge $(u,v)$. This scoring follows a feed-forward scheme:

\begin{equation}
Y_{uv}^{(t)} = \mathbf{v}^T \tanh(\mathbf{W}_{key}\mathbf{z}_{uv} + \mathbf{W}_{query}\mathbf{q}^{(t)})
\end{equation}
where $\mathbf{W}_{key}$ and $\mathbf{W}_{query}$ are trainable weight matrices, and $\mathbf{v}$ is a learnable parameter vector. The raw attention scores are first passed through a $\tanh$ activation and scaled by a learnable factor. To ensure validity, the raw scores are scaled by a factor $d$ and adjusted by a feasibility mask to prohibit invalid edges. The final probability distribution is then computed as: ${P}^{(t)} = \text{softmax}(\text{mask}(d \cdot Y^{(t)}))$. The next edge $\tau_t$ is sampled from this distribution during training or chosen deterministically via $\text{argmax}({P}^{(t)})$ during evaluation. Finally, the embedding of the selected edge is fed back to the LSTM for the next decoding step.

\vspace{\baselineskip}

\textbf{Entropy loss:} To encourage the policy to sample more diverse solutions, we add an entropy regularization term to the objective function.
Ideally, we would compute the full entropy of the entire sequence $p_\theta(\mathbf{\tau}|G)$, but this is computationally intractable. 
Therefore, we approximate the sequence entropy by summing the conditional entropies at each decoding step.

The entropy of the policy ($\mathcal{H}_{p_{\theta}}$) is used to quantify the randomness of the solution and serves to promote exploration during training. At each time step $t$, the Decoder provides the probability distribution ${P}^{(t)}$ over the candidate actions. The policy entropy is approximated by summing the entropy calculated from these distributions at each step:
\begin{equation}
\mathcal{H}_{p_{\theta}}= \sum_{t=1}^{\kappa} \mathcal{H}(p_{\theta}(\tau_t|G, \tau_{0:t-1}))
\end{equation}
The final entropy loss ($\mathcal{L}_{\text{entropy}}$) is defined by taking the average of the conditional entropies over the sequence length $\kappa$ for each instance, and subsequently averaging across the batch:

\begin{equation}
\mathcal{L}_{\text{entropy}} =  \frac{1}{\kappa} \sum_{t=1}^{\kappa} \mathcal{H}(p_{\theta}(\tau_t | G, \tau_{0:t-1})) 
\end{equation}


\textbf{Training: }
In the training phase, $G \in \Gamma$ where $\Gamma$ is the set of all instances in the training set and $\Gamma_i$ are the instances in batch $i$. The policy gradient $\nabla_{\theta} J(\theta|G)$ is formulated with two primary stabilizing terms: an Entropy Regularization term $\mathcal{H}_{p_{\theta}}$ for exploration, and a Central Self-Critic Baseline $b_{i}$ introduced in~\cite{ma2019combinatorial} for variance reduction. The term $r(s_{i,t}, a_{i,t})$ denotes the reward function, calculated as the negative sum of the edge weights associated with the chosen action given the current state. $\alpha$ is the weight parameter that controls randomness and B is the batch size. Let $\mathcal{I}$ and $\mathcal{A}$ denote the state space and action space, respectively. The state at time step $t$, denoted as $s_t \in \mathcal{I}$, is defined as the set of previously chosen edges. The action $a_t \in \mathcal{A}$ corresponds to the selection of the next edge.The policy gradient can be expressed as:

\begin{equation}
\nabla_{\theta} J(\theta |G) =  \frac{1}{B}\sum_{i=1}^{B}\Bigg[\sum_{t=1}^{\kappa}(r(s_{i,t}, a_{i,t}) - b_{i}) \cdot \nabla \log p_{\theta}(a_{i,t}|s_{i,t}) + \alpha\nabla \mathcal{H}_{p_{\theta}}\Bigg]
\end{equation}

\vspace{\baselineskip}
During training, the Central Self-Critic Baseline $b_{i}$ is expressed as:

\begin{equation}
\begin{aligned}
b_{i}
&=
\sum_{t=1}^{\kappa}
    r\!\left(\tilde{s}_{i,t}, \tilde{a}_{i,t}\right)
\\[3pt]
&\quad+
\frac{1}{B}
\sum_{j=1}^{B}
\sum_{t=1}^{\kappa}
\Big[
    r\!\left(s_{j,t}, a_{j,t}\right)
    -
    r\!\left(\tilde{s}_{j,t}, \tilde{a}_{j,t}\right)
\Big]
\end{aligned}
\label{eq:central-self-critic}
\end{equation}



Equation~\eqref{eq:central-self-critic} formulates the central self-critic baseline, denoted as $b_i$, which is designed to further reduce gradient variance compared to the standard self-critic method. The baseline consists of two components. The first term, $\sum_{t=1}^{\kappa} r(\tilde{s}_{i,t}, \tilde{a}_{i,t})$, represents the cumulative reward obtained by the greedy policy for the specific instance $i$. The second term acts as a batch-centered correction, calculating the average difference between the stochastic policy's reward and the greedy policy's reward across the entire batch $B$.

\section{Experiment}
In our experiments, we evaluated the multi-solution performance of our proposed methods, GPN-tree and GPN-TreeM, on standard TSP benchmarks. We conducted a comprehensive comparison against a diverse set of strong baselines, ranging from traditional approaches—such as the state-of-the-art Niching Memetic Algorithm~\cite{huang2019niching} and constructive Bi-Criteria Heuristic~\cite{gao2022obtaining}—to recent neural diversity solvers like RF-MA3S~\cite{li2025diversity}. The results demonstrate that our methods match the solution diversity of traditional solvers with significantly reduced inference time, while also surpassing existing neural methods in tour diversity.

\subsection{Experiment Settings}
\textbf{Datasets.} To comprehensively evaluate the proposed method, we conduct experiments on widely used practical instances from TSPLIB~\cite{reinelt1991tsplib}. Specifically, we selected instances of varying scales to test scalability: \textit{berlin52}, \textit{eil101}, \textit{rd400} , and \textit{rat783}, where the numeric suffix in each instance name corresponds to the number of cities in the graph.  Furthermore, we normalize the coordinates of the cities to the range $[0, 1]$. These datasets serve as standard benchmarks to assess both the optimality and the diversity of the generated tours across different problem sizes.

\textbf{Competitors.} We benchmark our framework against representative baselines from both traditional and existing neural categories:
\begin{itemize}
    \item \textbf{Traditional Heuristics:} We include the Niching Memetic Algorithm (NMA)~\cite{huang2019niching}, widely regarded as the state-of-the-art heuristic for Multi-Solution TSP. Additionally, we compare against the Bi-Criteria Heuristic~\cite{gao2022obtaining}. To allow for a direct comparison, we adapt its output—diverse spanning trees—into TSP tours using the same DFS traversal employed in our Method 1.
    
    \item \textbf{Neural Baseline:} We evaluate against RF-MA3S~\cite{li2025diversity}, a recent DRL-based solver specifically designed for diversity optimization. It employs an encoder-decoder architecture with adaptive active search to balance solution quality and diversity.
\end{itemize}

\textbf{Implementation.} All experiments are conducted on a machine equipped with an NVIDIA RTX 4090 GPU and an Intel Core i7-13700K CPU. Traditional heuristics are executed on the same hardware to ensure a fair comparison of computational efficiency.

\subsection{Hyperparameters and Evaluation Metrics}

\textbf{Our Setting.} The GPN models for sampling diverse spanning tree and perfect matching are both trained on 40 cities in random 2D settings. That is, the coordinates of each city is sampled via a uniform distribution from (0,0) to (1,1). Training is performed for 100 epochs with 1,000 steps per epoch and a batch size of 256 and 128, respectively. We optimize network parameters using the Adam optimizer with a constant learning rate of $5 \times 10^{-4}$. To systematically investigate the impact of exploration, we trained multiple variants parameterized by the entropy coefficient $\alpha$. While we explored a range of values during preliminary experiments, we report detailed results for the representative set $\alpha \in \{0, 1, 3, 7\}$. Validation is performed at the end of each epoch on generated instances to monitor convergence. The total training time are approximately 12.7 hours and 1.2 hours respectively per $\alpha$ setting.

\textbf{Baseline Settings.} To ensure a fair comparison, we align the hyperparameters of the baseline methods with the recommendations in their respective original papers.
\begin{itemize} 

    \item For the traditional Criteria Heuristic, we adopt the methodology in~\cite{gao2022obtaining}, which formulates the search for diverse spanning trees as a Constrained Minimum Spanning Tree (CMST) problem. In this framework, finding a distinct tree is treated as minimizing edge usage frequency (penalty) subject to a cost budget constraint. We implement their Lagrangian relaxation-based approximation to solve this CMST formulation, thereby generating diverse spanning trees. Consistent with our experimental design, we generate 25 trees for each relaxation factor $c$ within the range $[1.1, 5.0]$. Finally, to extend these diverse spanning trees to the TSP domain, we convert each tree into a valid tour via a DFS traversal.
    
    \item For the evolutionary baseline, NMA~\cite{huang2019niching}, we executed the algorithm with 8 distinct population sizes: $N_{pop} \in \{150, 200, 250, 300, 350, 400, 450, 500\}$. For each $N_{pop}$ setting, we repeatedly ran the algorithm to accumulate 125 solutions. We adopted specific hyperparameters for our experiments: a mutation rate $P_m=0.1$, a stagnation threshold of 60 for critical edge updates, and a niche radius refinement parameter of 2. Furthermore, the adaptive neighborhood strategy was constrained with minimum and maximum neighborhood sizes of $M_{\min}=4$ and $M_{\max}=6$, respectively.

    \item For the neural baseline, RF-MA3S~\cite{li2025diversity}, we trained the model on instances with 40 cities too. The training process took approximately 11 hours. Regarding RF-MA3S, we evaluated the version without AAS across all test instances. However, for the version with AAS, the evaluation was restricted to \textit{berlin52} and \textit{eil101} due to excessive inference time on the larger \textit{rd400} and \textit{rat783} instances.
    
\end{itemize}

\textbf{Inference and Dispersion Strategy.} Unlike methods that rely on computationally expensive active search during inference~\cite{li2025diversity}, our approach leverages the stochasticity of the trained policy. Except for RF-MA3S(without AAS), whose architecture does not support large-scale sampling, we generate a fixed pool of 1,000 candidate TSP tours for each test instance. To rigorously evaluate diversity, we follow the Dispersion Problem formulation defined in~\cite{gao2022obtaining}. From this pool of 1,000 solutions, we first filter to retain only those whose costs are within $c$ times the optimal cost (i.e., $Cost(\pi) \le c \times Cost(\pi_{opt})$). Finally, we select the $k$ most distinct tours from this filtered subset to maximize the diversity metric.

\textbf{Metrics.} We utilize the Jaccard Index to quantify the diversity of the solution set. A lower Jaccard value indicates higher diversity (dissimilarity). We report the average Jaccard Index along with its standard deviation across the selected subset of solutions for varying dispersion parameters $c \in \{2, 4, 8, 16\}$. This metric aligns with the diversity definitions used in recent literature~\cite{li2025diversity}, offering a robust assessment of the solution landscape coverage.

\subsection{Experimental Results and Analysis}

We present a comprehensive evaluation of our proposed frameworks, GPN-TreeM and GPN-Tree, benchmarking them against state-of-the-art traditional and neural baselines. Our experimental analysis is structured into two main parts: first, we investigate the impact of the entropy regularization coefficient $\alpha$ on the trade-off between solution quality and diversity; second, we conduct a comprehensive benchmarking on standard TSPLIB instances to jointly evaluate the structural diversity (Tables \ref{tab:jaccard_comparison_berlin52} to \ref{tab:jaccard_comparison_rat783}) and computational efficiency (Table \ref{tab:runtime_comparison}) of all methods. Regarding the diversity tables, it is important to note that if a column for a specific dispersion factor $c$ (e.g., $c=16$) is omitted, it indicates that the solution pool reached saturation at a stricter threshold. For clarity, missing entries for a dispersion factor c indicate that the solution pool saturated under that threshold, and “–” means that no feasible tour that satisfies the required conditions.

\begin{table}[ht]
    \centering
    \caption{Trade-off analysis under varying entropy parameter $\alpha$ settings, performed on instances containing 40 random cities uniformly distributed in the unit square $[0, 1]^2$. The values are presented as Mean $\pm$ Standard Deviation (the lower the better).}
    \label{tab:ablation_alpha}
    
    \footnotesize 
    \setlength{\tabcolsep}{0pt} 
    
    \begin{tabular*}{\linewidth}{@{\extracolsep{\fill}} c cc cc @{}}
        \toprule
        \multirow{2.5}{*}{\textbf{Lagrange Multiplier} $\alpha$} & \multicolumn{2}{c}{\textbf{Method 1: GPN-Tree}} & \multicolumn{2}{c}{\textbf{Method 2: GPN-TreeM}} \\
        \cmidrule{2-3} \cmidrule{4-5}
         & Avg. Cost  & Jaccard  & Avg. Cost  & Jaccard  \\
        \midrule
         0 & 6.32 $\pm$  0.27 & 0.47 $\pm$ 0.09 & 6.32 $\pm$ 0.54 & 0.43 $\pm$ 0.08 \\
         1 & 7.38 $\pm$ 0.40 & 0.16 $\pm$ 0.04 & 8.20 $\pm$ 0.75 & 0.15 $\pm$ 0.04 \\
         3 & 10.27 $\pm$ 0.83 & 0.08 $\pm$ 0.03 & 10.83 $\pm$ 0.99 & 0.08 $\pm$ 0.03 \\
         7 & 14.71 $\pm$ 1.35 & 0.04 $\pm$ 0.02 & 14.77 $\pm$ 1.26 & 0.04 $\pm$ 0.02 \\
        \bottomrule
    \end{tabular*}
\end{table}

\subsubsection{Results under Different Entropy Settings}
Table \ref{tab:ablation_alpha} illustrates the sensitivity of our model to the entropy coefficient $\alpha$, evaluated on the trade-off between solution quality (Average Tour Cost) and diversity (Average Jaccard Index). We varied $\alpha$ within the range $\{0, 1, 3, 7\}$. As $\alpha$ increases, we observe a consistent decrease in the Jaccard Index, indicating significantly improved solution diversity. However, this comes at the cost of increased tour lengths. Specifically, when $\alpha=0$, the model prioritizes optimality, yielding the lowest cost but high solution overlap. Conversely, a larger $\alpha$ introduces strong randomness, successfully minimizing the Jaccard Index but leading to suboptimal tour costs. This demonstrates that $\alpha$ serves as a controllable modulator to balance the exploration-exploitation trade-off according to practical requirements.

\begin{table}[htbp]
    \centering
    \caption{Solution diversity on the \textit{berlin52} instance. For each dispersion factor $c \in \{2, 4\}$, the top $\mathbf{k=30}$ most diverse solutions are selected from the cost-filtered pool. Values represent Mean $\pm$ Standard Deviation.}
    \label{tab:jaccard_comparison_berlin52}
    
    \footnotesize
    \begin{tabular*}{\linewidth}{@{\extracolsep{\fill}}lcc@{}}
        \toprule
        & \multicolumn{2}{c}{\textbf{Jaccard Index }} \\
        \cmidrule(lr){2-3}
        \textbf{Method} & $\mathbf{c=2}$ & $\mathbf{c=4}$ \\
        \midrule
        Bi-Criteria Heuristic & $ 0.08 \pm 0.03$ & $\mathbf{0.01 \pm 0.01}$ \\
        NMA & $0.09 \pm 0.03$ & $0.08 \pm 0.03$ \\
        RF-MA3S(without AAS) & - & $0.37 \pm 0.07$ \\
        RF-MA3S & $0.73 \pm 0.03$ & $0.73 \pm 0.07$ \\
        GPN-Tree (Ours) & $\mathbf{0.07 \pm 0.02}$ & $0.02 \pm 0.01$ \\
        GPN-TreeM (Ours) & $0.08 \pm 0.03$ & $0.02 \pm 0.01$ \\
        \bottomrule
    \end{tabular*}
\end{table}

\begin{table}[htbp]
    \centering
    \caption{Solution diversity on the \textit{eil101} instance. For each dispersion factor $c \in \{2, 4, 8\}$, the top $\mathbf{k=60}$ most diverse solutions are selected from the cost-filtered pool. Values represent Mean $\pm$ Standard Deviation.}
    \label{tab:jaccard_comparison_eil101}
    
    \footnotesize
    \begin{tabular*}{\linewidth}{@{\extracolsep{\fill}}lccc@{}}
        \toprule
        & \multicolumn{3}{c}{\textbf{Jaccard Index }} \\
        \cmidrule(lr){2-4}
        \textbf{Method} & $\mathbf{c=2}$ & $\mathbf{c=4}$ & $\mathbf{c=8}$ \\
        \midrule
        Bi-Criteria Heuristic & $0.08 \pm 0.03$ & $0.02 \pm 0.02$ & $\mathbf{0.01 \pm 0.01}$ \\
        NMA & $0.09 \pm 0.03$ & $0.09 \pm 0.02$ & $0.09 \pm 0.03$ \\
        RF-MA3S(without AAS) & - & - & $0.32 \pm 0.05$ \\
        RF-MA3S & $0.51 \pm 0.16$ & $0.51 \pm 0.16$ & $0.51 \pm 0.16$ \\
        GPN-Tree (Ours) & $0.07 \pm 0.02$ & $0.02 \pm 0.01$ & $0.01 \pm 0.01$ \\
        GPN-TreeM (Ours) & $\mathbf{0.07 \pm 0.02}$ & $\mathbf{0.02 \pm 0.01}$ & $0.01 \pm 0.01$ \\
        \bottomrule
    \end{tabular*}
\end{table}

\subsubsection{Result on TSPLIB}

\textbf{Diversity Assessment on Small and Medium Instances.} 
For the \textit{berlin52} and \textit{eil101} instances (Tables \ref{tab:jaccard_comparison_berlin52} and \ref{tab:jaccard_comparison_eil101}), our proposed methods demonstrate remarkable capability in generating diverse solutions. Specifically, at a dispersion factor of $c=4$, GPN-Tree achieves Jaccard scores of $0.0163$ and $0.0156$ respectively, which are comparable to the results of the constructive Bi-Criteria Heuristic. This indicates that our neural policy successfully learns to emulate the high-quality topological dispersion of deterministic constructive algorithms. Furthermore, both of our methods significantly outperform other neural baselines; for instance, on \textit{berlin52}, our Jaccard scores are orders of magnitude better than RF-MA3S.

\textbf{Scalability to Large Instances.} 
The advantages of our approach become even more pronounced on larger instances like \textit{rd400} and \textit{rat783}, as shown in Tables \ref{tab:jaccard_comparison_rd400} and \ref{tab:jaccard_comparison_rat783}. As the optimality gap relaxes to $c=16$, our methods consistently discover highly distinct tours. Notably, on \textit{rd400} with $c=16$, GPN-Tree achieves a Jaccard Index of 0.0042, surpassing the Bi-Criteria Heuristic. Similarly, on the largest instance \textit{rat783}, our method achieves the best diversity score of 0.0023, significantly outperforming the Bi-Criteria Heuristic. This suggests that in high-dimensional spaces, our stochastic sampling strategy effectively explores the long-tail distribution of the solution landscape.

\textbf{Computational Efficiency.}
Table \ref{tab:runtime_comparison} demonstrates the critical advantage of our proposed framework in terms of inference speed. Although the Bi-Criteria Heuristic has a theoretical complexity of $\mathcal{O}(N^3)$, it shows an empirical scaling of roughly $\mathcal{O}(N^{2.2})$. Such super-quadratic growth becomes prohibitive at larger scales; for instance, on \textit{rat783}, NMA requires over 1.4 days, while the Bi-Criteria method takes nearly 1.8 days.

Theoretically, the baseline RF-MA3S (without AAS) functions as a city-selection model, whereas our frameworks employ a sequential edge-selection process. While this edge-based approach entails a higher worst-case complexity of $\mathcal{O}(N^3)$, the practical runtime is significantly mitigated by GPU parallelism. The computations in our encoder and decoder are implemented as dense matrix operations executed in parallel. The empirical runtime scales near-linearly across the range of \textit{berlin52}, \textit{eil101}, and \textit{rd400}. For instance, while the problem size increases approximately $8\times$ from \textit{berlin52} to \textit{rd400}, the runtime grows by only $10.5\times$. Even extending to the largest instance \textit{rat783}, the scaling behavior remains consistent with the neural baseline.

Consequently, GPN-TreeM and GPN-Tree offer the most compelling trade-off: they match or exceed the diversity quality of Bi-Criteria heuristics while maintaining the high-throughput inference speed characteristic of neural solvers, even when trained on small datasets of size $N = 40$ with city coordinates uniformly distributed between 0 and 1.

\begin{table}[htbp]
    \centering
    \caption{Solution diversity on the \textit{rd400} instance. For each dispersion factor $c \in \{2, 4, 8, 16\}$, the top $\mathbf{k=240}$ most diverse solutions are selected from the cost-filtered pool. Values represent Mean $\pm$ Standard Deviation.}
    \label{tab:jaccard_comparison_rd400}
    \footnotesize 
    \setlength{\tabcolsep}{1pt} 
    \begin{tabular*}{\linewidth}{@{\extracolsep{\fill}}lcccc@{}}
        \toprule
        & \multicolumn{4}{c}{\textbf{Jaccard Index }} \\
        \cmidrule(lr){2-5}
        \textbf{Method} & $\mathbf{c=2}$ & $\mathbf{c=4}$ & $\mathbf{c=8}$ & $\mathbf{c=16}$ \\
        \midrule
        Bi-Criteria Heuristic & $0.178 \pm 0.110$ & $0.028 \pm 0.026$ & $0.016 \pm 0.027$ & $0.016 \pm 0.027$ \\
        NMA & $\mathbf{0.113 \pm 0.031}$ & $0.048 \pm 0.020$ & $0.047 \pm 0.020$ & $0.047 \pm 0.020$ \\
        RF-MA3S(without AAS) & - & - & - & $0.173 \pm 0.016 $\\
        GPN-Tree (Ours) & $0.154 \pm 0.018$ & $\mathbf{0.026 \pm 0.006}$ & $0.012 \pm 0.004$ & $\mathbf{0.004 \pm 0.002}$ \\
        GPN-TreeM (Ours) & $0.188 \pm 0.019$ & $0.067 \pm 0.057$ & $\mathbf{0.011 \pm 0.004}$ & $0.004 \pm 0.002$ \\
        \bottomrule
    \end{tabular*}
\end{table}

\begin{table}[htbp]
    \centering
    \caption{Solution diversity on the \textit{rat783} instance. For each dispersion factor $c \in \{2,4,8,16\}$, the top $\mathbf{k=480}$ most diverse solutions are selected from the cost-filtered pool. Values represent Mean $\pm$ Standard Deviation.}
    \label{tab:jaccard_comparison_rat783}
    \footnotesize
    \setlength{\tabcolsep}{1pt} 
    \begin{tabular*}{\linewidth}{@{\extracolsep{\fill}}lcccc@{}}
        \toprule
        & \multicolumn{4}{c}{\textbf{Jaccard Index }} \\
        \cmidrule(lr){2-5}
        \textbf{Method} & $\mathbf{c=2}$ & $\mathbf{c=4}$ & $\mathbf{c=8}$& $\mathbf{c=16}$ \\
        \midrule
        Bi-Criteria Heuristic & $\mathbf{0.189 \pm 0.114}$ & $\mathbf{0.040 \pm 0.040}$ & $\mathbf{0.017 \pm 0.031}$ & $0.017 \pm 0.031$ \\
        NMA & $\text{-}$ & $0.049 \pm 0.029$ & $0.040 \pm 0.016$ & $0.040 \pm 0.016$ \\
        RF-MA3S(without AAS) & - & - & - & $0.110 \pm 0.011$ \\
        GPN-Tree (Ours) & $\text{-}$ & $0.046 \pm 0.006$ & $0.020 \pm 0.015$ & $0.003 \pm 0.001$\\
        GPN-TreeM (Ours) & $\text{-}$ & $0.058 \pm 0.007$ & $0.022 \pm 0.020$ & $\mathbf{0.002 \pm 0.001}$ \\
        \bottomrule
    \end{tabular*}
\end{table}

\begin{table}[htbp]
    \centering
    \caption{Comparison of running times for generating 1,000 solutions on TSPLIB instances.}
    \label{tab:runtime_comparison}
    \footnotesize
    \setlength{\tabcolsep}{4pt}
    \begin{tabular*}{\linewidth}{@{\extracolsep{\fill}}lcccc@{}}
        \toprule
        \textbf{Method} & \textbf{berlin52} & \textbf{eil101} & \textbf{rd400} & \textbf{rat783} \\
        \midrule
        Bi-Criteria Heuristic & 7.2m & 26.4m & 9.8h & 1.8d \\
        NMA & 9.3m & 28.8m & 8.1h & 1.4d \\
        RF-MA3S(without AAS) & 1.6s & 2.3s & 18.3s & 1.7m \\
        RF-MA3S & 2.2m & 10.2m & $>5$d & $>5$d \\
        GPN-Tree (Ours) & \textbf{1.2m} & \textbf{2.6m} & 12.6m & 1.1h \\
        GPN-TreeM (Ours) & 1.5m & 3.0m & 15.3m & 1.2h \\
        \bottomrule
    \end{tabular*}
\end{table}

\section{Conclusion}

In this work, we tackle the challenge of generating a set of diverse and approximated TSP tours, which is a key requirement for applications ranging from fault-tolerant routing to adversarial security. We propose a framework consisting of graph pointer network and dispersion algorithm, developing two novel neural approaches. The first, GPN-Tree, leverages diverse spanning tree sampling followed by randomized DFS traversals. The second, GPN-TreeM, extends this foundation by adapting the Christofides algorithm via a deep network for diverse minimum matchings.

The experiments on standard TSPLIB benchmarks demonstrate that our proposed methods successfully bridge the gap between traditional heuristics and modern neural solvers. By combining entropy-driven generation with dispersion-based selection, our approaches attain solution diversity comparable to state-of-the-art constructive heuristics while delivering a computational speedup of several orders of magnitude. Notably, on large-scale instances, our methods reduce solution generation time from 1.8 days to roughly 1.2 hours. Moreover, we outperform existing neural baseline in terms of both tour lengths and diversity. Our work confirms that entropy regularization method effectively addresses the scalability limitations of current neural combinatorial optimization, establishing a novel framework for the D-TSP.


\bibliographystyle{plainurl}
\bibliography{reference}

\newpage        
\appendix      
\section{Additional Experimental Results} 
\label{app:extra_results}

In this appendix, we examine the inherent generation capabilities of different models across two standard benchmarks of varying scales: \textit{berlin52} and \textit{eil101}. Specifically, we analyze the performance \textbf{the ``Dispersion Problem and Greedy Selection'' stage}. This allows us to directly evaluate the raw solution quality and the intrinsic diversity captured by the stochastic policies of each model.

\subsection{Analysis of Distributions}
\begin{figure}[ht]
    \centering
    
    \begin{minipage}{0.48\linewidth} 
        \centering
        \includegraphics[width=\linewidth]{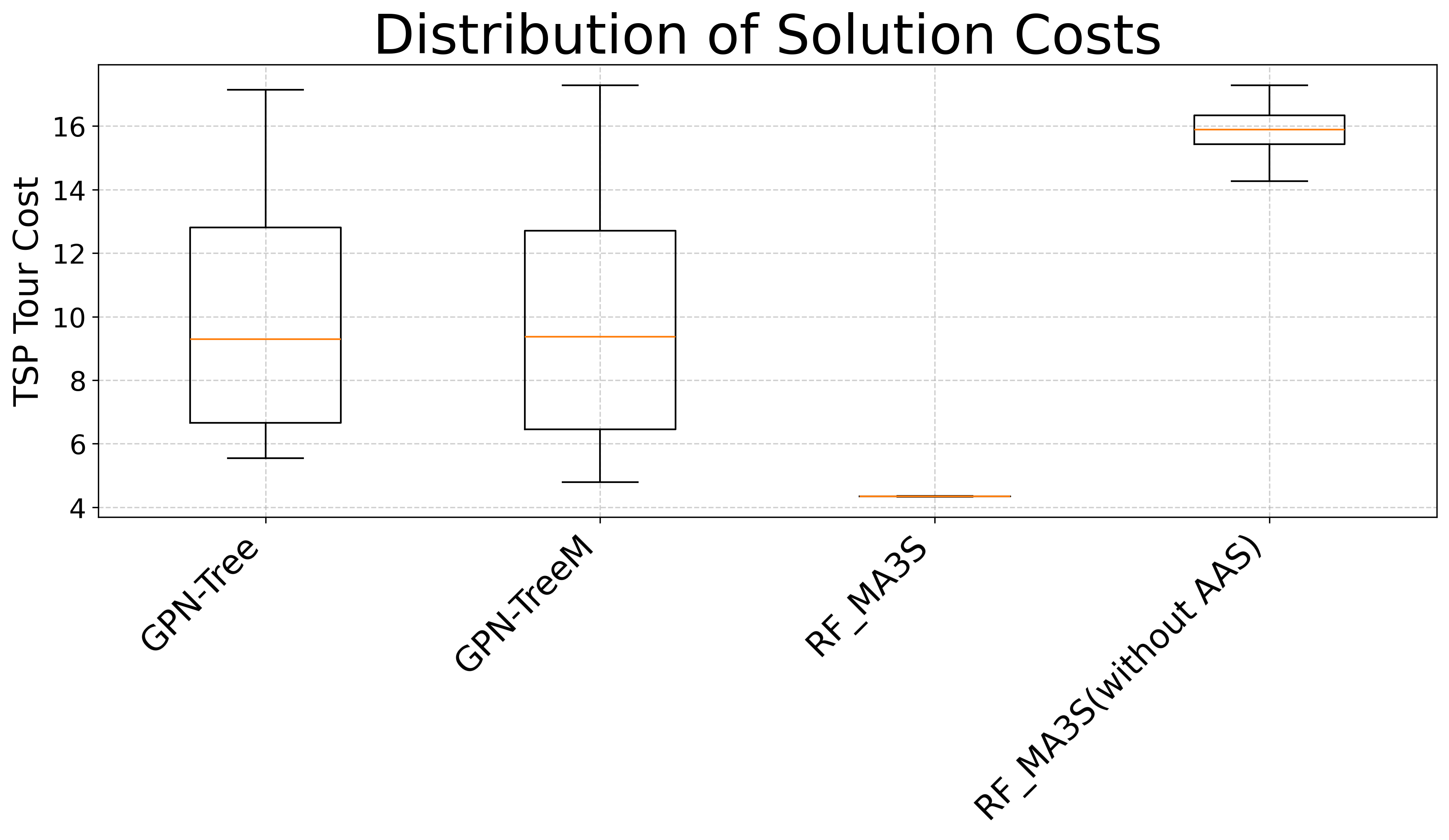}
        \centerline{(a) \textit{berlin52}: Cost Distribution.}
    \end{minipage}
    \hfill 
    \begin{minipage}{0.48\linewidth} 
        \centering
        \includegraphics[width=\linewidth]{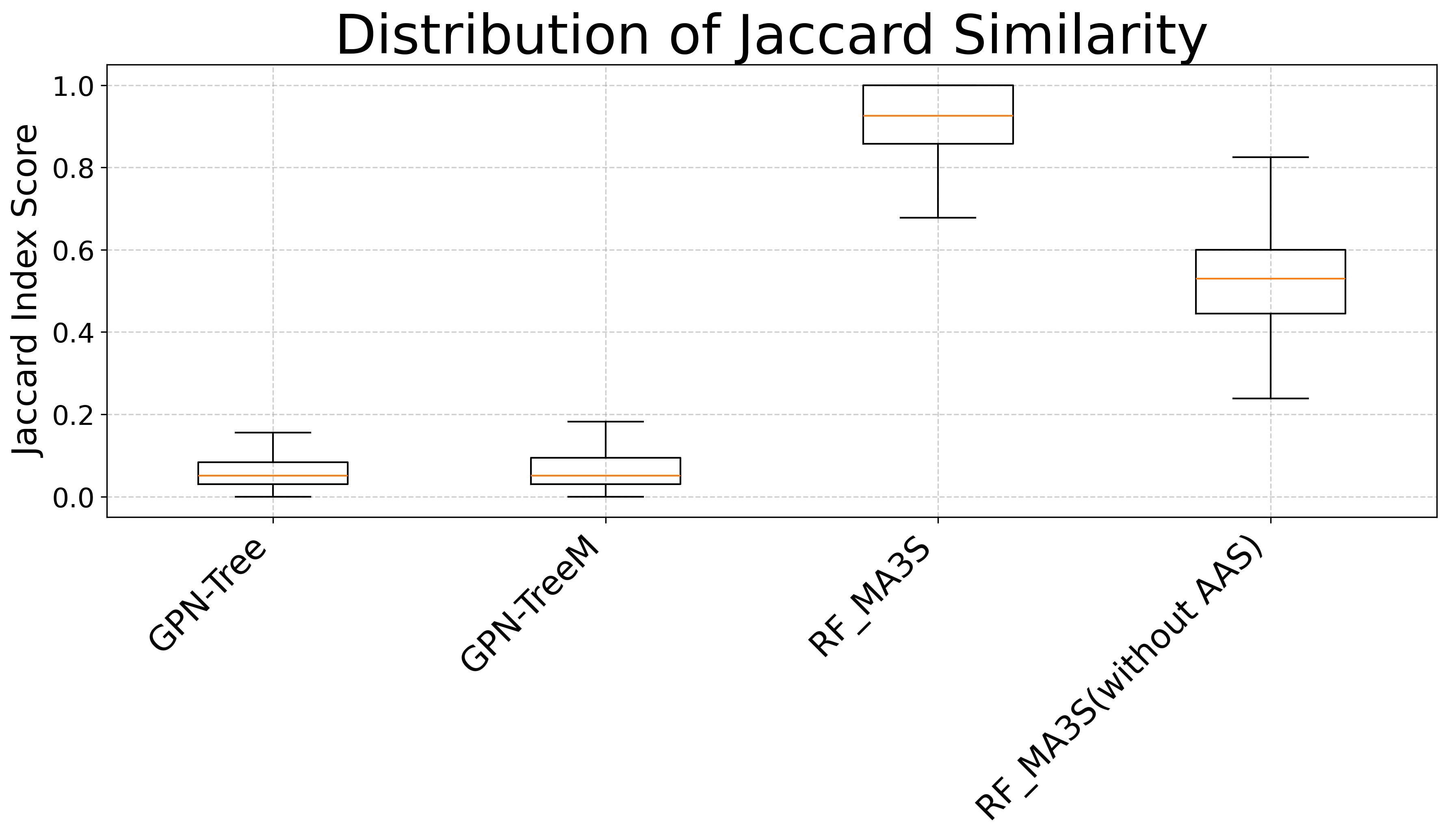}
        \centerline{(b) \textit{berlin52}: Jaccard Index Distribution.}
    \end{minipage}
    
    \vspace{0.5cm} 
    
    \begin{minipage}{0.48\linewidth} 
        \centering
        \includegraphics[width=\linewidth]{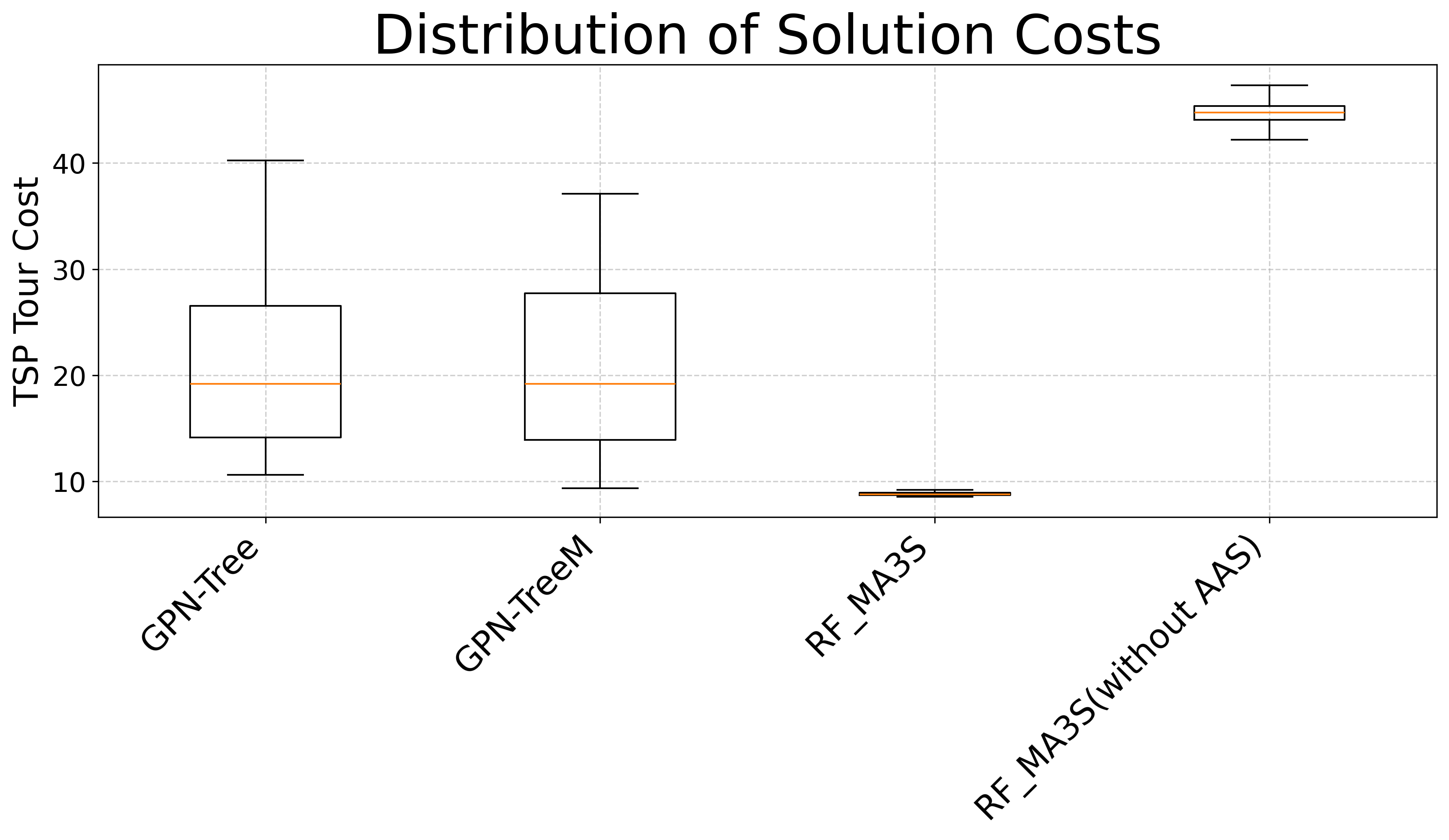}
        \centerline{(c) \textit{eil101}: Cost Distribution.}
    \end{minipage}
    \hfill 
    \begin{minipage}{0.48\linewidth} 
        \centering
        \includegraphics[width=\linewidth]{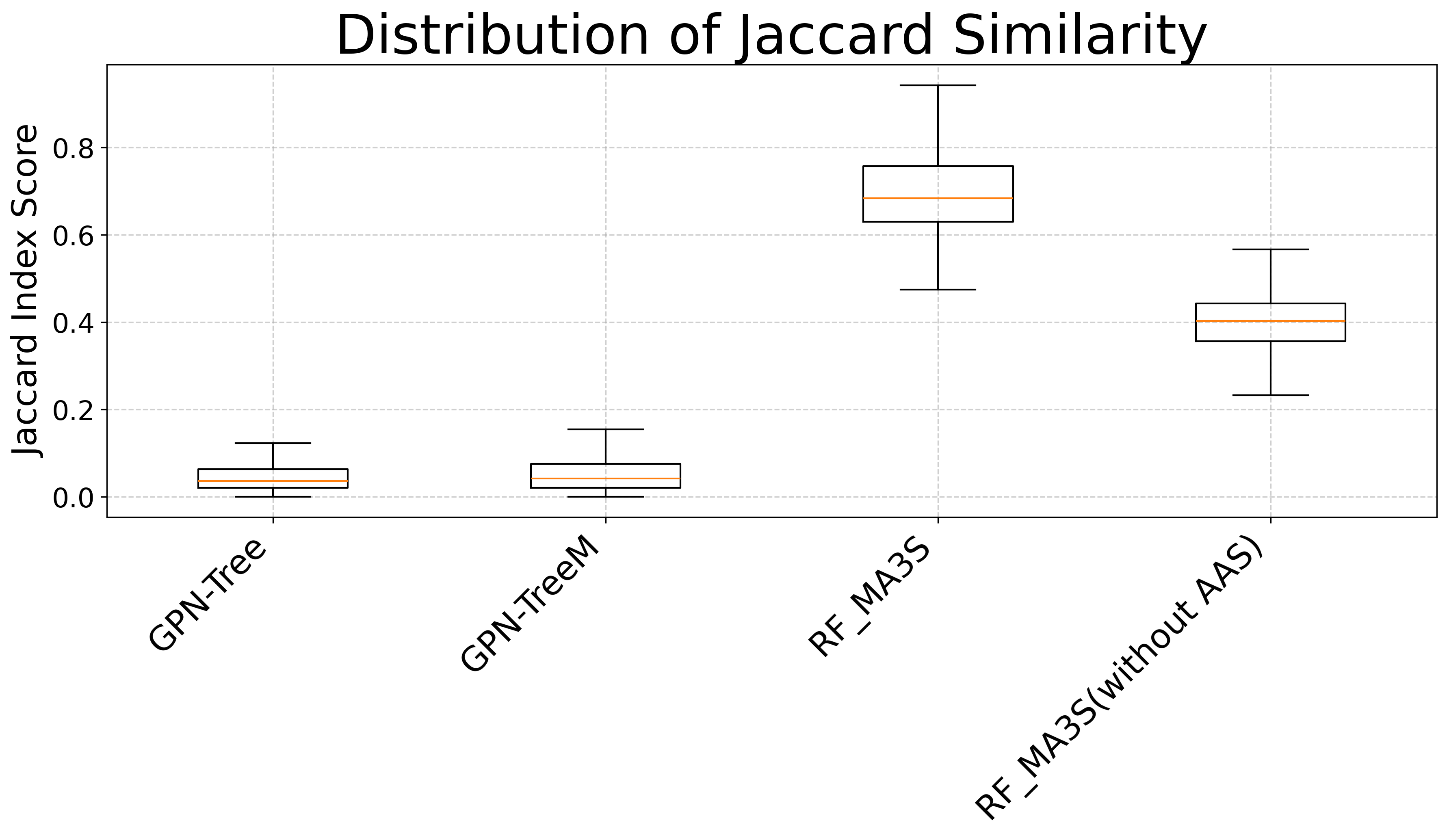}
        \centerline{(d) \textit{eil101}: Jaccard Index Distribution.}
    \end{minipage}
    
    \caption{Performance comparison across different scales. Top row: \textit{berlin52}; Bottom row: \textit{eil101}. The left column displays cost distributions, while the right column shows Jaccard index distributions.}
    \label{fig:distributions_combined} 
\end{figure}

As illustrated in Figure \ref{fig:distributions_combined}, we observe distinct patterns between our method and the baselines:
\begin{itemize}
    \item \textbf{Optimality (Left Column):} 
    The cost distributions (Figures \ref{fig:distributions_combined}a and \ref{fig:distributions_combined}c) show that our method  maintains a competitive solution quality. 
    
    \item \textbf{Diversity (Right Column):} 
    Crucially, the Jaccard index distributions (Figures \ref{fig:distributions_combined}b and \ref{fig:distributions_combined}d) reveal the advantage of our approach.  This confirms that our stochastic policy inherently captures a wider variety of structural modes even before the explicit dispersion selection step is applied.
\end{itemize}

\end{document}